\documentclass[%
 aip,
 amsmath,amssymb,
 reprint,%
]{revtex4-1}

\usepackage{graphicx}
\usepackage{dcolumn}
\usepackage{bm}
\usepackage{amsmath,amssymb,amsfonts}
\usepackage[utf8]{inputenc}
\usepackage[T1]{fontenc}
\usepackage{mathptmx}
\usepackage{color}
\usepackage{etoolbox}
\usepackage{stmaryrd}

\newcommand{\blue}[1]{\textcolor{black}{#1}}
\newcommand{\pfr}[2]{\ensuremath{\frac{\partial #1}{\partial #2}}}
\newcommand{\pfi}[2]{\ensuremath{{\partial #1}/{\partial #2}}}
\newcommand{\ep}{\varepsilon}

\newcommand{\vect}[1]{\mathbf{#1}}

\newcommand{\oline}[1]{\overline{#1}}

\DeclareMathAlphabet\mathbfcal{OMS}{cmsy}{b}{n}

\makeatletter
\def\@email#1#2{%
 \endgroup
 \patchcmd{\titleblock@produce}
  {\frontmatter@RRAPformat}
  {\frontmatter@RRAPformat{\produce@RRAP{*#1\href{mailto:#2}{#2}}}\frontmatter@RRAPformat}
  {}{}
}%
\makeatother
\begin{document}

\preprint{AIP/123-QED}

\title[Hydrodynamic theory of premixed flames under Darcy's law]{Hydrodynamic theory of premixed flames under Darcy's law}
\author{Prabakaran Rajamanickam}
\altaffiliation[Present address: ]{Department of Mathematics \& Statistics, University of Strathclyde, Glasgow G1 1XQ, UK}
\author{Joel Daou}%
 \email{joel.daou@manchester.ac.uk}
\affiliation{ 
Department of Manchester, University of Manchester, Manchester M13 9PL, UK
}%


\date{\today}

\begin{abstract}
This paper investigates the theoretical implications of applying Darcy's law to premixed flames, a topic of growing interest in research on flame propagation in porous media and confined geometries. A   multiple-scale analysis is carried out treating the flame as a hydrodynamic discontinuity in density, viscosity and permeability. The analysis accounts in particular  for the inner structure of the flame.  A simple model is derived allowing the original conservation equations to be replaced by Laplace's equation for pressure, applicable on both sides of the flame front, subject to specific conditions across the front. Such model is useful for investigating general problems under confinement including flame instabilities in porous media or Hele-Shaw channels. In this context,  two Markstein numbers are identified, for which explicit expressions   are provided.  In particular, our analysis reveals novel contributions to the local propagation speed arising from discontinuities in the tangential components of velocity and gravitational force, which are permissible  in Darcy's flows to leading order, but not in flows obeying Euler or Navier--Stokes equations.
\end{abstract}

\maketitle

\section{Introduction}
\label{sec:intro}

The study of flame propagation in porous media and narrow confined geometries, such as  Hele-Shaw cells, is an active research area \citep{al2019darrieus,dejoan2024effect,fernandez2018analysis,veiga2019experimental,veiga2020unexpected}. One key distinction between normal flames and flames in strongly confined media lies in their hydrodynamic behaviour. In porous media,  fluid dynamics is primarily governed by Darcy's law and so is the case in slender Hele-Shaw cells. Several recent numerical investigations have successfully employed Darcy's law to characterize flame propagation in Hele-Shaw cells, under the assumption of near-adiabatic walls~\citep{fernandez2018analysis,martinez2019role,rajamanickam2024effect,fernandez2019impact}.

A fundamental question remains however: What are the theoretical implications   of applying Darcy's law to premixed flames? A recent study by~\citet{daou2025hydrodynamic} addressed this question by treating the flame as a hydrodynamic discontinuity and investigating its instabilities. The study builds upon earlier research by~\citet{joulin1994influence}, Miroshnichenko \textit{et. al.}~\cite{miroshnichenko2020hydrodynamic}, which utilized the so-called Euler--Darcy model. Despite these advancements, a comprehensive theoretical description of premixed flames under Darcy's law, accounting for their internal structure, remains an open problem. This paper is dedicated to addressing this challenging problem.

The classical hydrodynamic theory of premixed flames, based on the Navier-Stokes equations, was developed in pioneering contributions by~\citet{ashinsky1988nonlinear,clavin1982effects,pelce1988influence,matalon1982flames,clavin1983premixed}. Notably, the latter two studies presented comprehensive theories that accounted for finite-amplitude flame wrinkling. Subsequent research has extended this theory to various contexts~\citep{clavin1983influence,garcia1984soret,clavin1983premixed,clavin1985effect,keller1994transient,matalon2003hydrodynamic,matalon2009multi,clavin2011curved,bechtold2024hydrodynamic}. 

In this paper, we embark on a theoretical analysis of premixed flames assumed to be governed by Darcy's law, an assumption which  is motivated by  asymptotic analyses in the narrow channel limit~\cite{fernandez2018analysis,rajamanickam2024effect,daou2025hydrodynamic}. One of our main aims is to understand the hydrodynamic aspects of confinement on flame propagation, which are poorly understood. To isolate and highlight these aspects, intimately related to Darcy's law, we consider a simplified configuration involving an equi-diffusive reacting mixture (unity Lewis number) and neglect heat-loss effects. As we shall see, our study will result in the derivation of a simplified hydrodynamic model for flame propagation, involving specific jump conditions across the flame front and explicit formulas for the so-called Markstein numbers, characterising the local propagation speed. The study will also reveal novel contributions to the local propagation speed arising from leading-order tangential discontinuities in the velocity and the gravitational force, which are not present in conventional flame theory.

\section{Problem formulation}
\label{sec:formulation}

\begin{figure}
\centering
\includegraphics[scale=0.57]{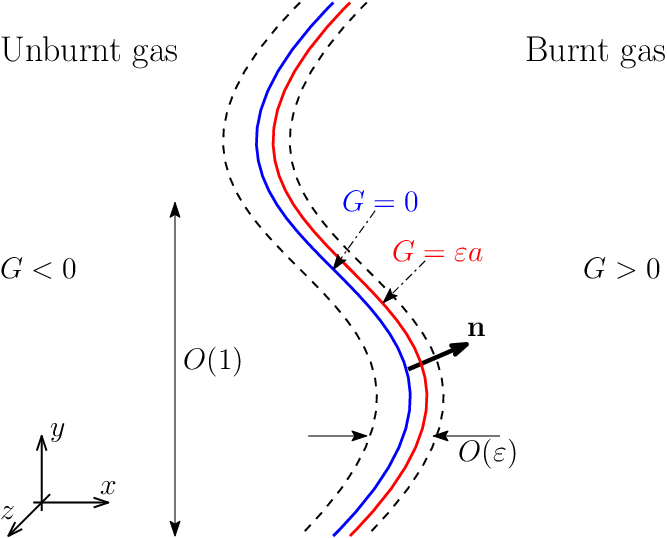}
\caption{Schematic illustration of a curved premixed flame propagating into a fresh mixture. The location of the flame-front corresponds to a level set $G=0$ of the field $G(\vect x,t)$; the location of the reaction sheet corresponds to the level set $G=\ep a$.} 
\label{fig:sch}
\end{figure}

Consider a premixed flame propagating through an unburnt gas mixture containing a deficient reactant. This mixture possesses constant values of  density $\rho_u$, viscosity $\mu_u$, permeability $\kappa_u$ and thermal diffusivity $D_u$.  Similarly, the burnt gas mixture behind the flame is characterised by constant properties, $\rho_b$, $\mu_b$, $\kappa_b$ and $D_b$. For Hele-Shaw channels, $\kappa_u=\kappa_b = h^2/12$, where $h$ is the channel width. We assume that the characteristic length scale, $L$, of flame wrinkling is significantly larger than the flame thickness $\delta_L=D_u/S_L$, where $S_L$ represents the planar, laminar flame speed \blue{with respect to the unburnt gas}. Our theoretical framework is based on a small expansion parameter, $\ep$, defined as
\begin{equation}
    \ep = \frac{\delta_L}{L} \ll 1.
\end{equation}
For convenience, we non-dimensionalise physical quantities using $L$ as the length scale, $L/S_L$ as the time scale, $S_L$ as the velocity scale and $\mu_u D_u/\ep \kappa_u$ as the pressure scale. As previously mentioned, the Lewis number of the reacting mixture will be assumed equal to one and heat loss effects will be ignored. Under these conditions, the mass fraction $Y$ of the reactant is simply related to the temperature $T$ by $Y/Y_u = 1-(T/T_u-1)/q$, where $q$ defines the flame temperature by the relation $T_b = T_u(1+q)$. All physical properties of the fluid are non-dimensionalised using their respective values on the unburnt gas mixture and the non-dimensional temperature $\theta$ is defined by $\theta=(T-T_u)/qT_u$ such that it approaches zero in the unburnt gas and unity in the burnt gas. A schematic illustration of the premixed flame, in non-dimensional units, is depicted in Fig.~\ref{fig:sch}.

Let us introduce the familiar $G$-equation~\citep{williams1985turbulent},
\begin{equation}
   \rho\left(\pfr{G}{t}+\vect v\cdot \nabla G\right) = \dot m |\nabla G|, \label{Geq}
\end{equation}
which involves $\dot m(\vect x,t)$, the  normal mass flux  crossing a given level set $G(\vect x,t)=$ const. The function $G(\vect x,t)$ defines for a given level set the local normal  unit vector $\vect n=\nabla G/|\nabla G|$  pointing towards the burnt gas. We identify the level $G=0$ to be the flame-front location from the viewpoint of the outer hydrodynamic zone. Furthermore, the reaction-sheet location will be identified with the level $G=\ep a$, where $a$ is some constant.

On each side of the reaction sheet $G\neq \ep a$, the following governing equations are assumed to hold,
\begin{align}
    \pfr{\rho}{t} + \nabla \cdot(\rho \vect v)& = 0,\\
    -\mu \vect v = \nabla p &- \rho \vect g,  \label{darcy}\\
    \rho \pfr{\theta}{t} + \rho \vect v\cdot \nabla \theta & =  \ep\nabla\cdot(\lambda \nabla \theta) , \label{theta}\\   
    \rho = \rho(\theta), \quad \mu =&\mu (\theta), \quad \lambda = \lambda(\theta),
\end{align}
where $\vect g$ is the non-dimensional gravity vector, whose magnitude, $|\vect g|=\rho_u g \kappa_u/\mu_u S_L$, measures the strength of the gravitational force. The function $\mu=\mu(\theta)$ is assumed to incorporate both fluid viscosity and permeability.

Across the reaction sheet (not the flame front), all physical variables satisfy certain jump conditions, which are are well established and are expressed readily in terms of a small-scale flame coordinate $\zeta$, which is defined by
\begin{equation} 
    \zeta  =  \frac{1}{\ep}(G-\ep a).   \label{smallcoord}
\end{equation}
At the reaction sheet, $\zeta=0$, we have
\begin{align}
  \llbracket \vect v \rrbracket=\llbracket p \rrbracket=\llbracket \theta \rrbracket=\lambda|\nabla G|\left\llbracket\pfr{\theta}{\zeta}\right\rrbracket+1=0, \label{sheetjump}
\end{align}
where $\llbracket\varphi\rrbracket\equiv \varphi |_{\zeta=0^+}-\varphi |_{\zeta=0^-}$.  In summary, $G=0$ (or $\zeta=-a$) corresponds to the flame front, i.e., the front location seen from the viewpoint of the outer hydrodynamic zone and $G=\ep a$ (or $\zeta=0$)  corresponds to the reaction sheet. Traditionally, one usually sets $a=0$, although it is important to recognise that the end results of the analysis will depend on $a$ as can be inferred from the analysis by~\citet{bechtold2001dependence} and as such we will not discard the constant parameter $a$.

\section{Multiple-scale analysis}

The asymptotic solution to the problem described above in the limit $\ep\to 0$ is carried out now using multiple-scale analysis, following closely~\citet{keller1994transient}\footnote{The analysis of \citet{keller1994transient} differs from others in the sense that the former works in a fixed Cartesian coordinate system $(\vect x,t)$ instead of the flame-fixed coordinate system.}. In our multiple-scale analysis, all physical variables, except $G$ (and therefore $\vect n$), are assumed to depend on both the large-scale hydrodynamic coordinates $(\vect x,t)$ and  the small-scale flame coordinate, $\zeta$. Any physical variable $\varphi$ is expanded in a power series in $\ep$:
\begin{equation}
    \varphi = \varphi_0(\zeta,\vect x,t) + \ep \varphi_1(\zeta,\vect x,t) + \cdots
\end{equation}
It is important to note that $G$ cannot be a function of $\zeta$, as $G$ itself defines $\zeta$~\eqref{smallcoord}. Moreover, we shall assume from the outset that the function $G(\vect x,t)$ is described with arbitrary accuracy in powers of $\ep$ and as a result, we do not expand the function $G$, following~\citet{clavin2011curved,clavin2016combustion}. This assumption does affect the intermediate steps of the analysis, but not the final uniformly-valid solution.

For physical variables depending also on the coordinate $\zeta$, the transformation rule for derivatives with respect to large-scale coordinates $(x^i,t)$ are given by
\begin{equation}
    \pfr{}{x^i}\mapsto \frac{1}{\ep} \pfr{G}{x^i}\pfr{}{\zeta} + \pfr{}{x^i}, \qquad  \pfr{}{t}\mapsto \frac{1}{\ep} \pfr{G}{t}\pfr{}{\zeta} + \pfr{}{t}.
\end{equation}
Any function, which is independent of $\zeta$, can be regarded as variables corresponding to outer hydrodynamic zone. Furthermore,  the continuity equation can be rewritten,  when combined with the $G$-equation, as
\begin{equation} \label{modcont}
 \pfr{\rho}{t}+\nabla\cdot(\rho\vect v) = -\frac{|\nabla G|}{\ep} \pfr{\dot m}{\zeta}       \,.
\end{equation}
This has a simple interpretation: the continuity equation based on the outer coordinates $(\vect x,t)$ has a non-zero source  term whenever the normal mass flux $\dot m$ varies with the inner coordinate $\zeta$.

\subsection{Structure of the locally planar flame}

At the leading order, we obtain
\begin{align} \label{leading}
    \pfr{\dot m_0}{\zeta} =    \pfr{p_0}{\zeta} =0, \quad 
    \frac{\dot m_0}{|\nabla G|} \pfr{\theta_0}{\zeta} = \pfr{}{\zeta}\left(\lambda_0\pfr{\theta_0}{\zeta}\right).
\end{align}
The solution, subject to the jump conditions~\eqref{sheetjump}, is given by
\begin{align} \label{leadingsol}
    p_0 = P_0(\vect x,,t),  \quad \dot m_0 = \dot M_0(\vect x,t) = 1,  \quad \theta_0 = \begin{cases}
        e^{\hat\zeta/|\nabla G|}, \,\, \zeta<0,\\
        1, \qquad \,\,\,\,\,  \zeta>0,
    \end{cases} 
\end{align}
where $\hat\zeta = \int_0^\zeta d\zeta/\lambda_0$. The outer functions $\dot M_0(\vect x,t)$ and $P_0(\vect x,t)$ are continuous across $G=0$. The continuity of $P_0(\vect x,t)$, which is specific to Darcy's law, will be shown later.

\subsection{Leading-order flow field}
At the first order, the momentum equation yield
\begin{equation}
    -\mu_0 \vect v_0 = \nabla P_0 - \rho_0 \vect g + \nabla G \pfr{p_1}{\zeta}.
\end{equation}
The pressure gradient $\pfi{p_1}{\zeta}$ can be eliminated by multiplying the equation vectorially with $\vect n$ to result in $-\mu_0 \vect v_0 \times \vect n = \nabla P_0 \times \vect n - \rho_0 \vect g \times \vect n$. Another cross product with $\vect n$ from the left yields the tangential component of $\vect v_0$, i.e., $\mathbfcal P\vect v_0=\vect n\times (\vect v_0\times\vect n)$, where $\mathbfcal P =\vect I -\vect n\vect n$ is the projection (matrix) operation of a vector onto the tangent surface. On the other hand, the normal component $\vect v_0\cdot\vect n$ can be determined from the leading-order $G$-equation $\rho_0 (\pfi{G}{t}+\vect v_0\cdot \nabla G) =  |\nabla G|$. Combining the two components, we obtain
\begin{equation} \label{v0Eq} 
    \vect v_0 = \left(\frac{1}{\rho_0} - \frac{1}{|\nabla G|}\pfr{G}{t}\right) \vect n - \frac{\mathbfcal P}{\mu_0}(\nabla P_0 - \rho_0 \vect g).
\end{equation}
It is convenient to  introduce an auxiliary outer function $\vect V_0(\vect x,t)$ defined for $G\neq 0$ by
\begin{align}
    \nabla \cdot \vect V_0=0, \quad -\oline{\mu}\,\vect V_0 = \nabla P_0 -\oline\rho\vect g, \label{outerflow0} \\ \oline\rho \frac{\tilde DG}{\tilde D t}\equiv \oline\rho \left(\pfr{G}{t}+\vect V_0 \cdot \nabla G\right) =  |\nabla G| \label{outerflow}
\end{align}
and involving the  (outer) constants $\oline\rho$ and $\oline\mu$ given  by 
\begin{equation}
    \oline\rho=\begin{cases}\begin{aligned}
        &   1,   \qquad\qquad\,\,\,\,\, G  <0,\\
       & \rho_f\equiv \rho(1),  \quad G>0,
    \end{aligned}\end{cases} \quad 
     \oline\mu=\begin{cases}\begin{aligned}
        &   1,   \qquad \qquad\,\,\,\,\,\, G<0,\\
       & \mu_f\equiv \mu(1),  \quad G>0.
    \end{aligned}\end{cases} \label{outerdensity}
\end{equation}
From the last two equations in~\eqref{outerflow0}-\eqref{outerflow}, it follows as in the derivation of~\eqref{v0Eq} that 
\begin{equation} \label{V0Eq}
    \vect V_0 = \left(\frac{1}{\oline \rho} - \frac{1}{|\nabla G|}\pfr{G}{t}\right) \vect n - \frac{\mathbfcal P}{\oline \mu}(\nabla P_0 - \oline \rho \vect g) \,,
\end{equation}
and therefore, on combining~\eqref{v0Eq} and~\eqref{V0Eq},  that
\begin{equation} 
    \vect v_0 -\vect V_0 = \frac{\oline\rho-\rho_0}{\oline\rho\rho_0}  \vect n - \mathbfcal P\left(\frac{\mu_0-\oline\mu}{\mu_0}\vect V_0 + \frac{\oline\rho-\rho_0}{\mu_0}\vect g\right) \,. \label{innerflow}
\end{equation}
This equation implies that $\vect v_0-\vect V_0$ vanishes exponentially as $\zeta\to- \infty$ and is identically zero for $\zeta>0$. Thus, the flow field $(\vect V_0,P_0)$, which is incompressible and obeys  Darcy's law  for $G \neq 0$  according to~\eqref{outerflow0}-\eqref{outerflow}, is indeed the outer flow to leading order.

\subsection{First correction to normal mass flux}
The continuity equation~\eqref{modcont} at the first order implies
\begin{align}
    |\nabla G|\pfr{\dot m_1}{\zeta} &= - \pfr{\rho_0}{t} - \nabla\cdot(\rho_0\vect v_0) \\ &= \frac{\tilde D }{\tilde D t}(\oline \rho-\rho_0)  - \nabla \cdot [\rho_0(\vect v_0 -\vect V_0)] \label{m1equation}
\end{align}
where the last expression follows from the previous  one, upon adding and subtracting  $\nabla \cdot (\rho_0 \vect V_0)$ and using the condition  $\nabla \cdot   \vect V_0 =0$ and the notation of~\eqref{outerflow}. We now integrate this equation from $\zeta=-\infty$ to an arbitrary location $\zeta$ in the preheat zone, noting that the integration can be commuted with the (outer) differential operators on the right side.  The integration is conveniently performed by changing the integration variable to $\theta_0= e^{\hat\zeta/|\nabla G|}$, as given by~\eqref{leadingsol}, so that $d\zeta = d\theta_0\, |\nabla G| \lambda_0/\theta_0 $. Carrying out the integration and using~\eqref{innerflow} we find
\begin{align}
    |\nabla G|[\dot m_1 - \dot M_1(\vect x,t)] = & \frac{\tilde D }{\tilde D t}(\hat{\mathcal I_1} |\nabla G|)  - \frac{1}{\oline\rho}\nabla \cdot(\hat{\mathcal I_1}\nabla G) \nonumber \\& + \nabla\cdot [|\nabla G|\mathbfcal P (\hat{\mathcal I_2} \vect V_0+\hat{\mathcal I_3} \vect g)] \label{m1}
\end{align}
 for $\zeta<0$, where $\dot M_1(\vect x,t)$ is the integration constant and 
\begin{align} \nonumber
    \hat{\mathcal I_1} (\theta_0)&=  \int_0^{\theta_0} \frac{\lambda_0}{\theta_0}(\oline\rho-\rho_0) d\theta_0, \\  \nonumber \hat{\mathcal I_2}(\theta_0) &= \int_0^{\theta_0} \frac{\rho_0\lambda_0}{\mu_0\theta_0}(\mu_0-\oline\mu)d\theta_0,\\ \nonumber
   \hat{\mathcal I_3}(\theta_0)& = \int_0^{\theta_0} \frac{\rho_0\lambda_0}{\mu_0\theta_0}(\oline\rho-\rho_0)d\theta_0.
\end{align}
The function $\dot m_1 -\dot M_1$ is seen to vanish exponentially as $\zeta\to-\infty$ (since $\theta_0 \to 0$ in the integrals) and identically for $\zeta>0$ (since the rhs of~\eqref{m1equation} is then zero). Thus, $\dot M_1(\vect x,t)$ corresponds to the first-order correction to the normal mass flux $\dot m$ in the outer region.
Evaluated at the location of the reaction sheet, $\zeta=0$  where $\theta_0 = 1$, the integrals $(\hat{\mathcal I_1},\hat{\mathcal I_2}, \hat{\mathcal I_3})$ become $(\mathcal I_1,\mathcal I_2,\mathcal I_3) \equiv (\hat{\mathcal I_1}(1),\hat{\mathcal I_2}(1), \hat{\mathcal I_3}(1))$ and come out of the derivatives in~\eqref{m1}. Dividing by  $|\nabla G|$ and simplifying using a few vector identities\footnote{Specifically, using the leading-order outer $G$-equation $\oline\rho\tilde DG/\tilde Dt=|\nabla G|$ given in~\eqref{outerflow}, we find following~\citet{keller1994transient} that $\tilde D|\nabla G|/\tilde Dt = (1/\oline\rho)  \vect n \cdot \nabla |\nabla G| -\vect n\vect n : \nabla \vect V_0 |\nabla G|$ and $\nabla^2 G=  \vect n \cdot \nabla |\nabla G| + |\nabla G|\nabla\cdot \vect n $, which simplifies the first two terms on the right side of~\eqref{m1}. To simplify the last term, we note that for an arbitrary vector $\vect A$,  $\mathbfcal P\vect A= \vect n\times(\vect A\times\vect n)$ and hence $\nabla\cdot [|\nabla G| \mathbfcal P \vect A] = - \nabla G \cdot [\nabla \times(\vect A\times \vect n)]$; dividing by $|\nabla G|$ we thus obtain $\nabla\cdot [|\nabla G| \mathbfcal P \vect A]/|\nabla G|= -\vect n\cdot\nabla\times(\vect A\times\vect n) =  - (\vect A\cdot\vect n)(\nabla\cdot\vect n) - \vect n\vect n:\nabla \vect A$, upon imposing the condition $\nabla\cdot\vect A=0$.}
we obtain for $\dot m_1-\dot M_1$ evaluated at $\zeta=0$ the expression
\begin{align}
   \dot m_1-\dot M_1   = & -  \left[\frac{\mathcal I_1}{\oline\rho}+ (\mathcal I_2\vect V_0+\mathcal I_3\vect g)\cdot\vect n\right]\nabla\cdot\vect n \nonumber \\ &- (\mathcal I_1 +\mathcal I_2)  \vect n\vect n:\nabla \vect V_0  \,.
\end{align}

We next examine the temperature equation  at the first order  which implies that
\begin{align}
    &|\nabla G|^2 \pfr{}{\zeta} \left(\lambda_0 \pfr{\theta_1}{\zeta} + \lambda_1 \pfr{\theta_0}{\zeta}\right) - \dot m_0|\nabla G| \pfr{\theta_1}{\zeta} \nonumber \\ & = |\nabla G|\pfr{(\dot m_1 \theta_0)}{\zeta} + \pfr{(\rho_0\theta_0)}{t} + \nabla \cdot (\rho_0 \vect v_0 \theta_0) \nonumber \\  &- \nabla^2G\lambda_0 \pfr{\theta_0}{\zeta} 
    - \nabla G\cdot \left[\pfr{}{\zeta}\left(\lambda_0 \nabla\theta_0\right) + \nabla \left(\lambda_0 \pfr{\theta_0}{\zeta}\right)\right] . \label{temp1}
\end{align}
In the burnt gas,  $\zeta>0$, the solution for $\theta_1$ is simply $\theta_1=0$. In the unburnt gas, it is sufficient to  integrate equation~\eqref{temp1} from $\zeta=-\infty$ to $\zeta=0^-$  to determine $\dot M_1$.  The integration is subject to the  requirement that $\theta_1$ and its gradient vanish as $\zeta\to-\infty$ and the conditions~\eqref{sheetjump} at the reaction sheet which imply that
\begin{equation}
     \theta_1= 0 \,, \quad |\nabla G| \left(\lambda_0  \pfr{\theta_1}{\zeta}+ \lambda_1 \pfr{\theta_0}{\zeta}\right) = 0 \quad \text{at} \quad \zeta=0^- \,.
\end{equation} 
Performing the integration, we obtain after some simplifications\footnote{The simplifications   use   the relation $\lambda_0\pfi{\theta_0}{\zeta}=\theta_0/|\nabla G|$ and the fact that  $\nabla\theta_0=0$ at $\zeta=0$ which follow from the expression of $\theta_0$ in~\eqref{leadingsol}.} the expression
\begin{align}
   \dot M_1  = &   \left[\frac{\mathcal J_1}{\oline\rho}+ (\mathcal J_2\vect V_0+\mathcal J_3\vect g)\cdot\vect n\right]\nabla\cdot\vect n  \nonumber \\ & + (\mathcal J_1 +\mathcal J_2)  \vect n\vect n:\nabla \vect V_0 \label{tempflux}
\end{align}
where
\begin{align}
    \mathcal J_1 &= \int_0^1 \frac{\lambda_0}{\theta_0}[\oline\rho-\rho_0(1-\theta_0)] d\theta_0 \,, \\
    \mathcal J_2 &= \int_0^1 \frac{\rho_0\lambda_0}{\mu_0\theta_0}(\mu_0-\oline\mu)(1-\theta_0)  d\theta_0 \,, \\
  \mathcal J_3 &= \int_0^{1} \frac{\rho_0\lambda_0}{\mu_0\theta_0}(\oline\rho-\rho_0)(1-\theta_0) d\theta_0 \,.
\end{align}
It is worth pointing out at this stage that the influence of  the constant parameter $a$ appearing in equation~\eqref{smallcoord} and in Fig.~\ref{fig:sch} is buried in these integrals. This is so since  $\oline\rho$ and $\oline\mu$  have jumps at $\zeta=-a$ or $G=0$, which corresponds to an isotherm contour  $\theta_0=\theta_*$ where 
$\theta_* \equiv e^{\hat\zeta_*/|\nabla G|}$ with  $\hat\zeta_* =-\int_{-a}^0 d\zeta/\lambda_0$. For illustration, the expression for  $\mathcal J_1$ is given by
\begin{align}
    \mathcal J_1=  & \int_0^{\theta_*} \frac{\lambda_0}{\theta_0}[1-\rho_0(1-\theta_0)]d\theta_0 \nonumber \\ &+ \int_{\theta_*}^{1} \frac{\lambda_0}{\theta_0}[\rho_f-\rho_0(1-\theta_0)]d\theta_0 \,,
\end{align}
and similar expressions can be written for $\mathcal J_2$ and $\mathcal J_3$.   To summarise, it is is convenient to view  $\theta_*$ as a prescribed parameter, determining $a$ and allowing to define the flame front as the iso-temperature surface $\theta=\theta_*$. This observation has been emphasised in the recent works by~\citet{giannakopoulos2015consistent,giannakopoulos2019consumption}. 

\subsection{Continuity of the pressure field}
The continuity of the pressure field, follows directly, from the Darcy's law $\nabla p = -\mu \vect v+\rho \vect g$. Provided the  terms on the right-hand side  experience at most  finite jumps across the flame, as it is the case, integration of this equation across the flame front  shows that $p$ is continuous across the flame. This statement holds true at all orders of $\ep$.

\section{Summary of  results}

At this stage, we are able to describe the flame in a Darcy's flow as a hydrodynamic interface across which specific jump conditions, correct to order $\ep$,  must be satisfied. To this end let us write  $\vect V = \vect V_0 + \ep \vect V_1$, $P=P_0 + \ep P_1$, $\dot M=\dot M_0 + \ep \dot M_1$ and drop the overbars for $\oline\rho$ and $\oline\mu$. Then the problem is governed on each side of the flame front, $G \neq 0$, by the equations 
\begin{equation}
    \nabla \cdot \vect V = 0, \quad -\mu \vect V = \nabla P -  \rho \vect g \quad \Rightarrow \quad \nabla^2P = 0 \,.  \label{finalproblem}
\end{equation}
The corresponding jump conditions to be satisfied at  the flame front, $G=0$,  are 
 \begin{equation}
    \llbracket \rho(\vect V-\vect U)\cdot\vect n\rrbracket = 0, \qquad \llbracket P\rrbracket =0 \,,  \label{interfacedarcy}
\end{equation}
where 
$\llbracket\varphi\rrbracket\equiv \varphi |_{G=0^+}-\varphi |_{G=0^-}$ and $\vect U \cdot \vect n = - (\pfi{G}{t})/|\nabla G|$ is the normal frame-front velocity. In addition the problem is constrained by the kinematic condition
 \begin{equation} \rho \left(\pfr{G}{t}+\vect V\cdot \nabla G\right)=\dot M |\nabla G|   \label{kinematic}
\end{equation}
characterising the propagation of the flame front. This condition is to  be applied either at $G=0^-$ (unburnt gas side where $\rho=1$) or  at $G=0^+$ (burnt gas side where $\rho=\rho_f$). It involves the normal mass flux  $\dot M$   given by   
\begin{equation}
    \dot M = 1 + \ep (\mathcal M_c \nabla\cdot \vect n + \mathcal M_s \vect n\vect n:\nabla\vect V) \label{m1final}
\end{equation}
and the two Markstein numbers 
\begin{equation}
    \mathcal M_c = \frac{\mathcal J_1}{ \rho}+ (\mathcal J_2\vect V+\mathcal J_3\vect g)\cdot\vect n \,, \quad \mathcal M_s = \mathcal J_1 + \mathcal J_2 \,. \label{markstein}
\end{equation}
The integrals $\mathcal J_1$, $\mathcal J_2$ and $\mathcal J_3$ appearing in the Markstein numbers depend on the choice for the flame-front location within the inner zone. As mentioned earlier, the  location  of the flame front  can be specified by selecting an
iso-temperature surface $\theta=\theta_*$ with  $0<\theta^*<1$. \blue{It is worth nothing here that the selection of the optimal $\theta_*$ when comparing quantitatively with numerical or experimental data,  seems to be a delicate matter, as discussed by~\citet{giannakopoulos2015consistent} within the common framework of Navier--Stokes flows. In the present framework of Darcy's flows, similar dedicated numerical studies in prototypical configurations are needed in  future investigations to assess the quantitative influence of the choice of $\theta_*$.}

In summary, the original problem  governed by the equations of Section~\ref{sec:formulation} has been reduced to  the simpler problem of solving Laplace's equation $\nabla^2P=0$ for $G\neq 0$ subject to  the conditions~\eqref{interfacedarcy}-\eqref{kinematic}.

A key novel result of this study corresponds to the  formulas~\eqref{markstein} for the Markstein numbers  which contain  new terms that are specific to Darcy's law. These formulas should be compared with the corresponding formulas~\citep{clavin2016combustion} based on Navier--Stokes equations which, when applied  under the same assumptions used here, imply that
\begin{equation}
    \mathcal M_c = \frac{\mathcal J_1}{\rho} \,, \quad \mathcal M_s = \mathcal J_1 .
\end{equation}
It is worth noting that the new terms in~\eqref{markstein} can be attributed to the Darcy's law allowing the presence of discontinuities to leading order in the tangential velocity across the flame. Such discontinuities arise either due to viscosity jumps, $\llbracket \mu \rrbracket \neq 0$, or  to jumps in the gravitation term, $\llbracket \rho  g_t\rrbracket  \neq 0$ where the subscript $t$ denotes tangential components. This can be seen from the jump condition  $\llbracket \mu V_t\rrbracket=\llbracket \rho  g_t \rrbracket $ which follows readily from an integration of the tangential component of Darcy's equation across the flame and use of the continuity of pressure, $\llbracket P  \rrbracket= 0$.  Such discontinuities of the tangential component of velocity across the flame do not occur (at least to leading order) for flames in a flow obeying Euler or Navier--Stokes equations, see e.g.,~\citet{matalon1982flames}. Interestingly, the curvature  Markstein number $\mathcal M_c$ in~\eqref{markstein} depends not only on the fluid physio-chemical properties, but also on the normal components of the flow velocity and gravity.

 The equations~\eqref{finalproblem}-\eqref{kinematic} are versatile, applicable to various problems of interest, \blue{involving both numerical and theoretical analysis.} In particular, they include explicit expressions for the Markstein numbers, which were to date unavailable, but which are valuable when studying flame propagation 
 and stability under confinement.  Such expressions 
 are useful, e.g., to complement our recent work~\citep{daou2025hydrodynamic} which explored the stability of a planar premixed flame propagating against a uniformly moving fresh mixture, accounting for the presence of  gravity, represented by vector $\vect g$ pointing in the direction of flame propagation. We derived in this case a dispersion relation linking the perturbation growth rate ($s$) to the transverse wavenumber magnitude ($k$), given by
\begin{equation}
    s = \frac{ak-bk^2}{1+ck},   \nonumber
\end{equation}
where
\begin{align}
    a = \frac{r-1}{1+m} + \frac{1-m}{1+m}\mathcal V - \frac{r-1}{1+m}\frac{m}{r}|\vect g|, \nonumber \quad &b  = \frac{r+m}{1+m}\mathcal M_c +  a \mathcal M_s, \\ &c = \frac{r-1}{1+m}\mathcal M_s \,.\nonumber
\end{align}
Here, $k$ and $s$ are measured in units of $\delta_L^{-1}$ and $S_L/\delta_L$,  $r=1/\rho_f$ is the unburnt-to-burnt gas density ratio, $m= 1/\mu_f$ is the viscosity ratio and $\mathcal V$ is the speed of the oncoming flow of fresh mixture, measured with $S_L$. The three terms in the expression for $a$ correspond, respectively, to Darrieus--Landau, Saffman--Taylor and Rayleigh--Taylor instabilities. The oncoming flow opposes flame propagation when $\mathcal V>0$ and aids flame propagation when $\mathcal V<0$. Similarly, $|\vect g|>0$ corresponds to downward flame propagation and $|\vect g|<0$ to upward flame propagation. More detailed analyses and implications of the dispersion relation can be found in~\citet{daou2025hydrodynamic}.

\blue{To close this paper, we mention worthwhile extensions of this work in future investigations. The first natural extension is to account for non-unit values of the Lewis numbers, as typically encountered in applications. In principle, this is straightforward to carry out despite the lengthy algebraic manipulations involved, as classically done within the so-called near-equidiffusive approximation~\cite{ashinsky1988nonlinear}. Another important issue to address is the influence of heat losses, which is a significant aspect to consider when describing the effect of confinement on flames. This complicating factor has been purposely sidelined in this paper in order to clarify the poorly understood hydrodynamic aspect associated with flame propagation in a Darcy's flow.  The effect of including heat losses in our model is definitely worth investigating, since it is anticipated to  modify the Markstein numbers, as can be inferred from similar studies~\cite{clavin1985effect,keller1994transient,matalon2009multi}.}

\begin{acknowledgments}
This work was supported by the UK EPSRC through grant EP/V004840/1  and Grant No. APP39756. The authors are grateful to Professor Paul Clavin for insightful comments on the paper.
\end{acknowledgments}

 \vspace*{0.5cm}
\textbf{Data availability statement:}
The data that support the findings of this study are available within the article.

\bibliography{aipsamp}

\end{document}